\documentclass[a4paper,fleqn,usenatbib]{mnras}
\usepackage{graphicx}
\usepackage[T1]{fontenc}
\usepackage{ae,aecompl}
\usepackage{amsmath}	
\usepackage{amssymb}	
\usepackage{bm}

\newfont{\caps}{cmcsc10}

\title[Negative Dynamical Friction]{Negative Dynamical Friction on compact objects moving through dense gas}

\author[A.\ Gruzinov, Y.\ Levin, C.~D.~Matzner]{
Andrei Gruzinov$^{1}$\thanks{ag92@nyu.edu}, Yuri Levin$^{2,3,4}$\thanks{yl3470@columbia.edu}, Christopher D.~Matzner$^5$\thanks{matzner@astro.utoronto.ca}
\\
$^{1}$Physics Department, New York University, 726 Broadway, NY
10003, USA
\\
$^{2}$Center for Theoretical Physics, Department of Physics,
Columbia University, New York, NY 10027, USA
\\
$^{3}$Center for Computational Astrophysics,
Flatiron Institute, New York, NY 10010, USA
\\
$^{4}$School of Physics and Astronomy,
Monash University, VIC 3800, Australia
\\
$^{5}$Department of Astronomy and Astrophysics,
University of Toronto, 50 St George St.,  Rm. 101  Toronto, ON   M5S 3H4   Canada
}

\begin{document}
\label{firstpage}
\pagerange{\pageref{firstpage}--\pageref{lastpage}}

\maketitle

\begin{abstract}
An over-dense wake is created by a gravitating object moving through a gaseous medium, and this wake pulls back on the object and slows  it down. This is conventional dynamical friction in a gaseous medium. We argue that if the object drives a sufficiently powerful outflow, the wake is destroyed and instead an extended under-dense region is created behind the object. In this case the overall gravitational force is applied in the direction of the object's motion, producing a negative dynamical friction (NDF).
Black holes in dense gas drive powerful outflows and may experience the NDF, although extensive numerical work is probably needed to demonstrate or refute this conclusively. NDF may be important for stellar-mass black holes and neutron stars inside ``common envelopes'' in  binary systems,  for stellar mass black holes inside AGN discs, or for massive black holes growing through super-Eddington accretion in early Universe. 


\end{abstract}


\section{Introduction}
\label{sec:intro}
\noindent

A massive object moving through a large cloud of lighter particles or gas, experiences dynamical friction \citep{cha, Ostriker, edg}  We specify our discussion to gas for the rest of the paper. It is easiest to visualize the physical origin of the friction in the frame of the moving object (Fig.~1): after passing the object, the gas streams curve towards the object's line of motion and encounter through a shock wave the overdense region that extends as a tail downstream. The gravitational attraction of the overdensity and the object results in the force on the object that is opposite to its motion.
This is the dynamical friction force.

Consider now a situation in which the object is the source of a powerful wind (Fig.~2). If the wind is able to overpower gravity, it creates a bow shock beyond the object's Bondi radius. As a consequence the material is pushed out away from the object's line of motion, creating an underdensity behind the object. In this paper we show that if the wind is fast enough, the underdensity is so extensive that the object experiences an overall gravitational force from the gas that is co-directed with its motion. Thus in this case the dynamical friction is negative. The object will accelerate!

In this paper we present a conceptual study of this effect. We note that a similar effect was identified in a paper of \cite{bogdanovic}, who considered a radiative feedback from an accreting black hole onto the medium. They found that the radiative feedback can destroy the wake and bring dynamical friction to the value close to zero. In fact some of their simulations showed acceleration, which was however orders of magnitude smaller than the original friction and was of no practical consequence. Also of considerable interest is the strong and robust  negative friction that has been found in a pioneering exploration by \cite{Masset} and \cite{MV},
for hot planetesimals moving through a dense gas, e.g.~in the protostellar disc.  There the under-density is caused by a thermal feedback from the moving object. The authors studied both an idealized case of homogeneous gas, as well as a more realistic case of such planetesimal embedded in a shearing disc. Here we focus our attention on compact objects in a dense gas, where a super-Eddington accretion is known to drive powerful outflows. In the context of accreting black holes or neutron stars, this effect may result in a strong acceleration that can have a substantial astrophysical significance.

The plan of our paper is as follows. In Section $2$, we consider an object with a simple spherical wind without gravity, 
and show that the dynamical friction in this case can be negative if the wind is sufficiently fast.
In Section $3$, we show that dynamical friction can {\it in principle} be negative for moving accreting objects, i.e.~we show that conservation laws allow negative total friction (hydrodynamic force plus dynamical friction). In Section $4$, we sketch possibly realistic scenarios for super-Eddington accretion onto black holes that result in a negative dynamical friction (NDF). For black holes in stellar envelopes \citep{mcl} and quasar disks \citep{art, levin03, levin} the NDF effect might be important. This could have major implication for the formation of compact binaries through a common 
envelope, as well as for compact objects embedded in a dense AGN disc. NDF could also be potentially important for intermadiate-mass BHs accreting at a super-Eddington rate in the early Universe, although other frictional effects may dominate.  In Section $5$, we briefly discuss these possible astrophysical consequences of the NDF. Throughout we emphasiza that
our results while suggestive, are inconclusive. To demonstrate, or refute the importance of NDF, one needs to perform realistic numerical experiments with non-spherical outflows and a non-homogeneous spacial distribution of gas that is appropriate for stellar envelopes or discs. 
\begin{figure}
\includegraphics[width=0.5\textwidth]{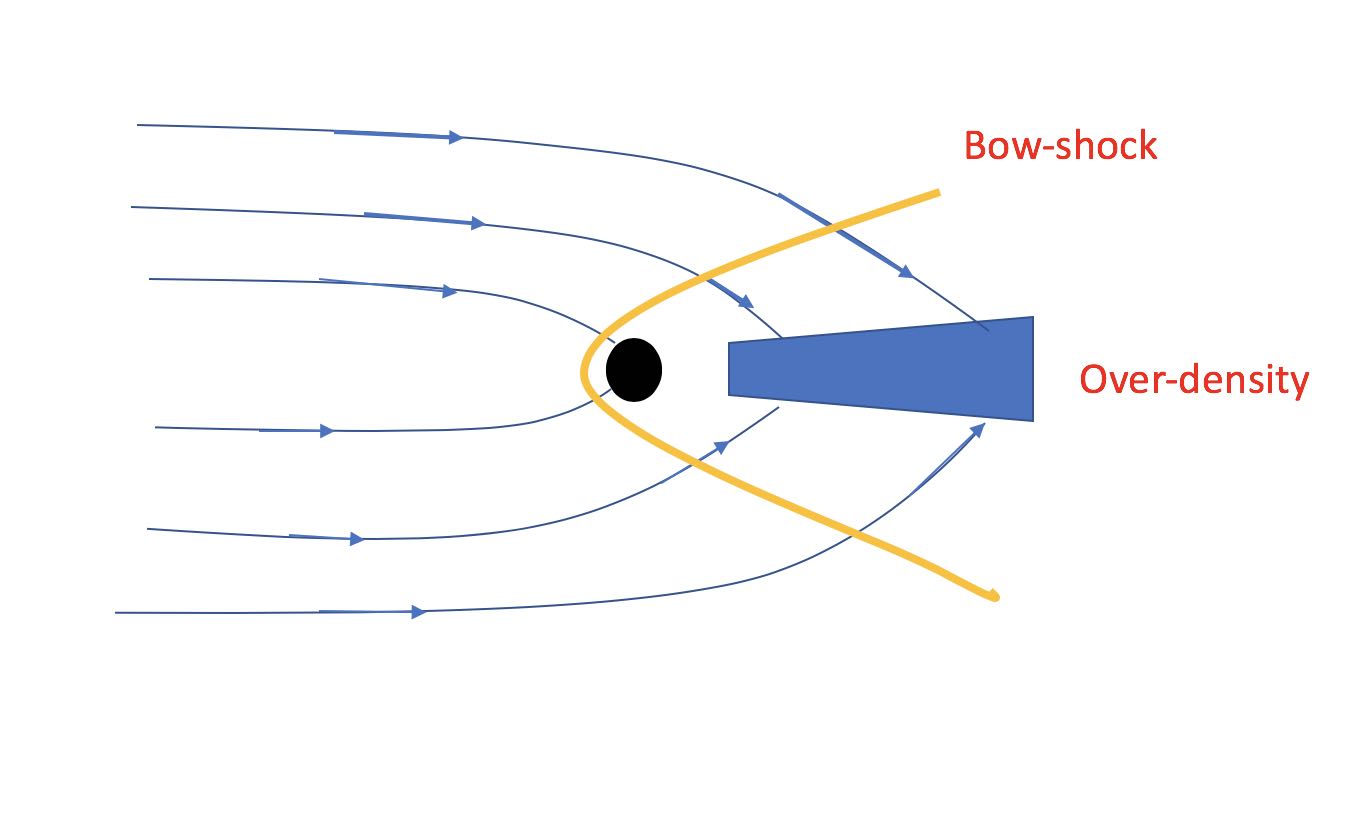}
\caption{Schematic view of Bondi-Hoyle accretion, as seen in the frame of reference of the moving object. The over-density creates  gravitational pull on the object, in the direction opposite to its velocity. This is conventional dynamical friction in a gaseous medium.} 
\label{fig:bowshock1}
\end{figure}

\begin{figure}
\includegraphics[width=0.5\textwidth]{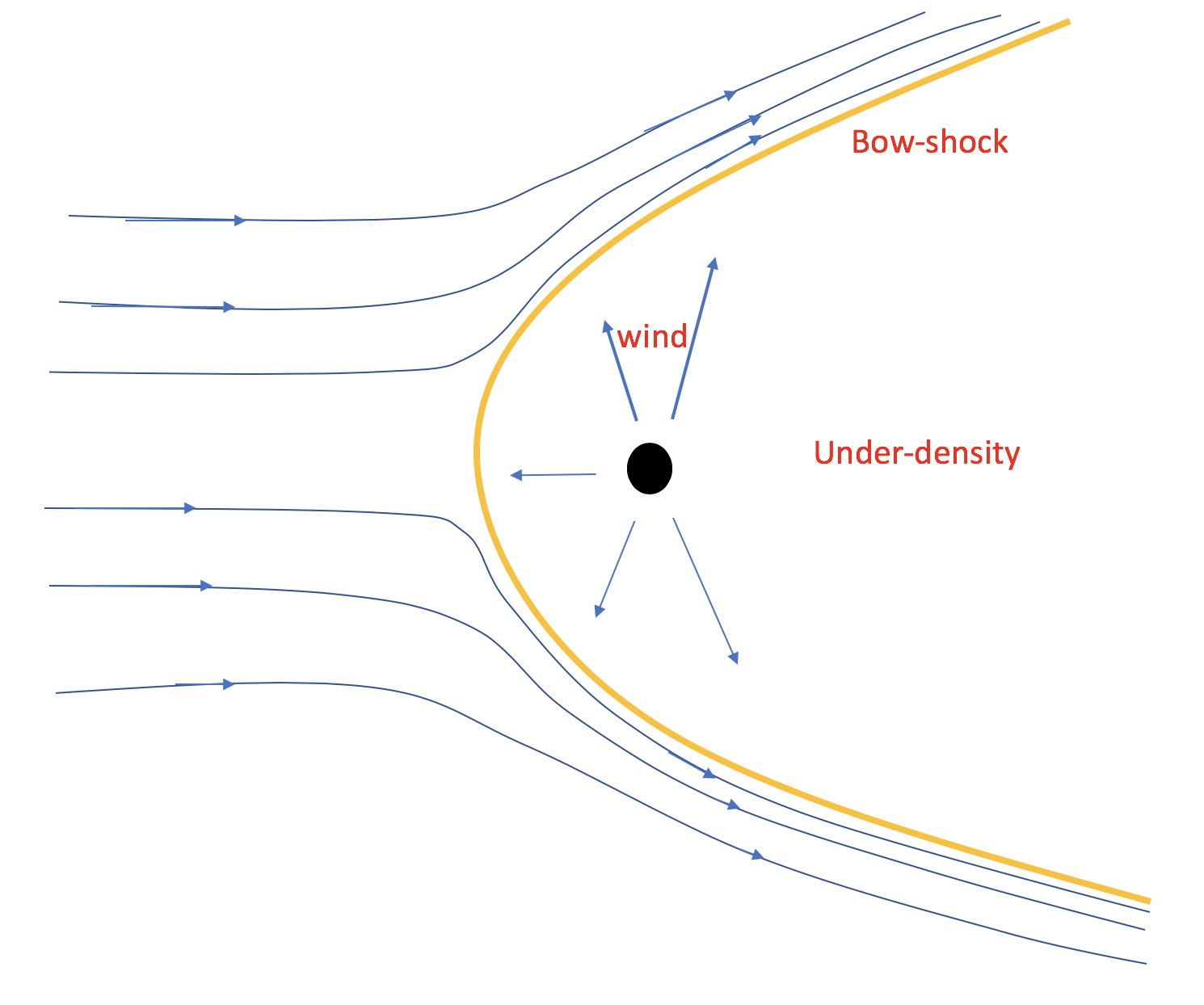}
\caption{Schematic view of the fluid flow in the frame of reference of the moving body with strong outflow. When the bow shock is pushed beyond the Bondi radius, it creates an under-density behind the object and the resulting gravitational pull is in the forward direction. The object accelerates.} 
\label{fig:bowshock2}
\end{figure}

\section{NDF from a spherical wind bow shock}
\label{sec:bow}
\noindent

Negative dynamical friction from a spherical stellar wind bow shock is easy to compute (in a certain approximation), but the effect is too small to be of any practical importance. We nevertheless carefully compute it here because: (i) rigorous results of the section prove that NDF is possible, (ii) we will apply the results of this section to accreting stars in a dense medium, where NDF might be of consequence.

Negative dynamical friction  from the  wind bow shock is an intuitively plausible result. If the wind emanating from the object is fast, it creates a large bubble of rarefied gas around the object. If the object is moving, the low-density bubble is asymmetrical, stretched behind the object; see Fig.~2. This gives a net gravitational force in the direction of motion -- the negative dynamical friction. 

We will see that this picture is in fact not accurate, because the gas compressed by the bow shock pulls the object back with about the same force as the bubble pulling it forward. One must really calculate the force in order to predict its sign. We describe the simplest possible calculation below.

Consider a spherical wind bow shock in the zero-temperature approximation \citep{bkk, wil}. The wind is characterized by the mass loss rate $\dot{M}$, and the velocity $V_w$. It emanates from an object moving with velocity $V_\star$ through a uniform cold ISM of density $\rho$. We consider the problem in the object's frame, where the emanating wind is assumed to be spherically symmetrical. The zero-temperature approximation means that when the cold  wind collides with the cold ISM headwind, the generated thermal energy is instantaneously radiated. 

The problem has only one dimensionless parameter
\begin{equation}
u\equiv \frac{V_\star}{V_w},
\end{equation} 
and we will work in the units
\begin{equation}
\rho=1,~~V_w=1,~~R_0\equiv \sqrt{\frac{\dot{M}V_w}{4\pi \rho V_\star ^2}}=1,
\end{equation} 
where $R_0$ is the standoff distance of the shock. 

\cite{wil} shows that, with the ISM velocity $u$ along negative $z$ and the star at the origin of coordinates, the bow shock is at the surface 
\begin{equation}
R_s=\csc \theta\sqrt{3(1-\theta\cot\theta)}
\end{equation} 
in spherical coordinates $(R,\theta,\phi)$, or
\begin{equation}
r_s=\sqrt{3(1-\theta\cot\theta)}
\end{equation} 
in cylindrical coordinates $(r,\phi,z)$. The surface density of the shock is 
\begin{equation}\label{eq:sig} 
\sigma =\frac{\left( 2u(1-\cos\theta)+r_s^2 \right) ^2}{2r_s\left( (\theta-\sin\theta \cos\theta)^2+(r_s^2-\sin ^2\theta)^2\right) ^{1/2}}.
\end{equation} 
The volume density of the gas is 
\begin{equation}\label{eq:rho} 
\rho =
  \begin{cases}
    1,      & R>R_s\\
    \frac{u^2}{R^2},  & R<R_s
  \end{cases}
\end{equation} 

Knowing the density, we calculate the gravitational force on the star of mass $M$ located at $R=0$. By symmetry, the force is along $z$. The force from the volume density is 
\begin{equation}
F_1=GM\int\limits_0^\infty R^2dR\int\limits_0^\pi 2\pi \sin\theta d\theta ~\cos\theta ~\frac{\rho}{R^2}.
\end{equation} 
With $\rho$ from Eq.(\ref{eq:rho}), after a regularization,
\begin{equation}
F_1=-2\pi GM\int\limits_0^\pi d\theta ~\sin\theta \cos\theta \left( R_s+\frac{u^2}{R_s}\right).
\end{equation} 
For future use, we rewrite this as
\begin{equation}
F_1=-2\pi GM\int\limits_0^\pi d\theta ~\cos\theta ~r_s\left( 1+\frac{u^2\sin^2\theta}{r_s^2}\right).
\end{equation} 

The force from the surface density (from the shock) is
\begin{equation}
F_2=GM\int\limits_0^\pi d\theta \left( \left( \frac{dr_s}{d\theta}\right)^2+\left( \frac{dz_s}{d\theta}\right)^2\right) ^{1/2}~2\pi r_s~\cos\theta ~\frac{\sigma}{R_s^2}.
\end{equation} 
With $\sigma$ from Eq.(\ref{eq:sig}),
\begin{equation}
F_2=\frac{3}{2}\pi GM\int\limits_0^\pi d\theta ~\cos\theta ~r_s\left( 1+\frac{2u(1-\cos\theta)}{r_s^2}\right)^2.
\end{equation} 

The net force, $F=F_1+F_2$, is
\begin{equation}
\begin{aligned}
F = & \pi GM\int\limits_0^\pi d\theta ~\cos\theta ~r_s\\
    & \left( \frac{3}{2}\left( 1+\frac{2u(1-\cos\theta)}{r_s^2}\right)^2
-2\left( 1+\frac{u^2\sin^2\theta}{r_s^2}\right) \right).
\end{aligned}
\end{equation}

For a slowly moving star, $u\ll 1$, we get negative dynamical friction -- the force is positive, along the velocity of the star:
\begin{equation}
F = -\frac{1}{2}\pi GM\int\limits_0^\pi d\theta ~\cos\theta ~r_s\approx 8.18~GM,~~~u\ll 1.
\end{equation}
For a fast star, $u\gg 1$, we get normal dynamical friction, the force is antiparallel to the velocity of the star:
\begin{equation}
\begin{aligned}
F = & 2\pi GMu^2\int\limits_0^\pi d\theta ~\cos\left( \frac{3(1-\cos\theta)^2}{r_s^3}-\frac{\sin^2\theta}{r_s}\right) \\
    & \approx -0.975~GMu^2,~~~u\gg 1.
\end{aligned}
\end{equation}
Numerical calculation also gives 
\begin{equation}
F = 0,~~~u\approx 1.71.
\end{equation}

Restoring dimensions, we summarize the results as
\begin{equation}\label{eq:fric} 
F =
  \begin{cases}
    2.31~GM\sqrt{\dot{M}\rho V_w}~V_\star^{-1},      & V_\star\ll V_w \\
    0,                                             & V_\star\approx 1.71~V_w \\
    -0.275~GM\sqrt{\frac{\dot{M}\rho}{V_w^3}}~V_\star ,      & V_\star\gg V_w
  \end{cases}
\end{equation} 
The result obviously does not apply for too small $V_\star$, when the ISM temperature, or the gravitational force on the ISM, or the finite ISM size must be included. 

NDF on a realistic windy star appears to be of no practical importance. To maximize the velocity change of a windy star, we take the maximal possible product of the force $F$ and the lifetime of the star $t$. Since $Ft\propto \sqrt{\dot{M}}t$, and $\dot{M}t\lesssim M$, the best we can do is to take a star which loses its entire mass in a Hubble time. Then a solar mass star, with initial velocity $V_\star= 10$ km/s, with high-velocity wind $V_w= 1000$ km/s, in a dense ISM $n=10~{\rm cm}^{-3}$ will be accelerated, after loosing all of its mass in a Hubble time, to $V_\star\approx 11$ km/s -- which is not a meaningful result, as far as we can see.

\section{Force on a moving accreting star: proof of principle}
\label{sec:negstar}
\noindent

Here we argue that the net friction force on an accreting star moving in a dense medium can be negative: the net force is in the direction of the velocity of the star. The effect is large enough to be of possible practical importance. 

Let us start with the following question of principle -- do conservation laws allow negative friction? Assume, uncritically, just to demonstrate the ideas, that in a dense medium standard Bondi-Hoyle-Littleton accretion regime still applies. To order of magnitude, this accretion regime can be described as follows. A supersonically moving star, velocity $V_\star$, in a medium of density $\rho$, accretes at a rate
\begin{equation}
\dot{M}_B\sim R_B^2\rho V_\star, ~~~R_B\sim \frac{GM}{V_\star^2},
\end{equation}
where we have introduced the Bondi-Hoyle radius $R_B$. The friction force on the star is positive, with hydrodynamic and dynamical friction forces making similar contributions:
\begin{equation}
F_B\sim \dot{M}_BV_\star\sim GM\rho R_B
\end{equation}

Energy conservation allows ejection of a fraction of the accreted mass, call it $\dot{M}_w$, with an arbitrarily high velocity $V_w$ (assumed non-relativistic, just for simplicity), provided $\dot{M}_BV_e^2> \dot{M}_wV_w^2$, where $V_e$ is the escape velocity from the surface of the star. For simplicity, assume that the mass is ejected along two oppositely directed 
{cones} 
perpendicular to the velocity of the star $V_\star$. The {wind cones} 
will pierce the accretion flow and terminate into two lobes, spreading into a wake in the downstream, at a distance of order the standoff distance of the wind bow shock 
\begin{equation}
R_w\sim \sqrt{\frac{\dot{M}_wV_w}{\rho V_\star^2}}\sim R_B\sqrt{\frac{\dot{M}_wV_w}{\dot{M}_BV_\star}}\gg R_B,
\end{equation}
where we have assumed 
\begin{equation}
\dot{M}_wV_w\gg \dot{M}_BV_\star,
\end{equation}
which is allowed by the constraint $\dot{M}_BV_e^2>\dot{M}_wV_w^2$. 

Since by assumption $V_w\gg V_\star$, we must use the first line of Eq.(\ref{eq:fric}) and estimate the NDF force as 
\begin{equation}
F_{\rm NDF}\sim \frac{GM}{V_\star}\sqrt{\dot{M}_w\rho V_w}~
\sim \sqrt{\dot{M}_BV_\star~\dot{M}_wV_w}
\gg F_B.
\end{equation}
The net force is a negative friction.  

{This result is intuitive, if we consider that a sheet at distance $R$ has  column density $\sigma\sim R\rho$ and gravity $\sim G \sigma\sim G R \rho$.  Because the wind shock lies farther from the star, its forward gravity exceeds the negative gravity of the wake.}

 Setting $\dot M_w\sim \dot M_B$ and allowing $F_{\rm NDF}$ act over a time $t$, it can accelerate the star to the speed 
\begin{equation} 
V_\star \sim \left( \frac{G^2 M \rho t}{V_w^3}\right)^{2/7} V_w 
\end{equation} 
at which $F_{\rm NDF}t = M_\star V_\star$.  

{In the estimates above we have used results from the axisymmetric wind shock and accretion flows, although our problem is not axisymmetric -- in fact it cannot be, as the accretion radius is closer to the star than the wind shock.  In a real problem, the angular dependence of the wind must introduce geometric factors: for instance, the accretion rate will be suppressed because part of the material within $R_B$ is ejected by wind; the wind output will be suppressed by a similar factor,  and the shape of the wind shock will no longer be a simple, cometary structure.  These effects may very well be time dependent.  However, they probably introduce only order-unity corrections to $F_{\rm NDF}/F_B$ unless the wind is very narrow.}

Note that a rocket force, which might exist if the wind itself is non-symmetrical, could conceivably be even larger  than $F_{\rm NDF}$: 
\begin{equation}
F_r\sim {\varepsilon} \dot{M}_wV_w 
\gg \varepsilon F_B
\end{equation}
{where $\varepsilon$ is the effective 
-- which must either develop within the wind's sonic or Alfv\'en point, or arise from a complex interaction with the inflow.}
This suggests that only very careful numerical simulations can properly describe the net force on an windy, accreting, moving star. 

Standard Bondi friction is known to be important in dense media, for example it leads to an inspiral during the common envelope phase. Then the much larger NDF (and, if present, rocket forces) must be carefully considered in any serious attempt to describe these accretion flows.  

As explained in the next section, we do not think that the estimates of this section count as a serious attempt to describe accretion flow in a dense medium. But these estimates do serve to show that NDF can, in principle, beat the standard friction.

\section{Force on a moving accreting black hole}
\label{sec:neg}
\noindent

Consider now a stellar mass black hole moving through a dense gas, like a common envelope of an interacting binary \citep{post,mcl}, or an accretion disc in a galactic nucleus \citep{art,levin}.
What is common for these situations, is that the Bondi-Hoyle accretion rate exceeds the Eddington limit by many orders of magnitude.

It is known from both observational and theoretical studies, that super-Eddington accretion onto black holes features  massive outflows. The observational golden standard for a super-Eddington accretion is the galactic micro-quasar SS433 (see Fabrika 2004 for a review). A consensus among the observers is that this source is powered by accretion onto $\sim 10 M_{\odot}$ black hole, fed through the inner Lagrange point from the Roche lobe of a more massive companion (e.g., Blundell et al.~2008, Cherepaschuk et al.~2019) at a rate of $\sim 10^{-4} M_{\odot}/\hbox{yr}$. As this is three orders of magnitude greater than the Eddington accretion rate, most of this material is ejected as a fast $\sim 1500\hbox{km}/\hbox{s}$ optically thick wind. 
This is in broad agreement with the expectations in \cite{shak},
who argued that for a thin disc with super-Eddington accretion rate a massive outflow should develop around the radius where the disc becomes quasi-spherical (i.e., $h\sim R$); see also \cite{beg}
and \cite{pout}.
This is also in agreement with general arguments that the outflows should always develop if the binding energy of the accreted gas is unable to be efficiently radiated \citep{shv,bla,qua,pan,gru}.
The system also features a mildly relativistic $\sim 0.3\hbox{c}$ jet. By measuring
its power and assuming that it equals $\sim 0.1 \dot{M}_{\rm BH} c^2$, one can estimate that the rate of mass increase of the black hole $\dot{M}_{\rm BH}$ is only several times greater than the Eddington accretion rate. In this work we concentrate on the feedback from the wind and neglect potentially very important feedback from the relativistic jet.

The strong outflows have also been observed in simulations. Over the past decade, numerical 
experiments on super-Eddington black hole accretion have become possible \citep{jiang,sad}.
While impressive,
the simulations are not yet able to model the flow over the required range of radii from outside of the photon-trapping radius to the black hole horizon. Nonetheless, both simulations observe development of a super-Eddington non-relativistic outflow that helps collimate the mildly relativistic jet, in a qualitative agreement with the observational data on SS433.

Let us now imagine a scenario in which the black hole is surrounded by a rotating accretion disc that is fed by a stream of the oncoming gas in the black hole's frame of reference; see Fig.~3.  If there was no feedback from the wind emanating from the disc, and if the disc radius was significantly smaller than the Bondi radius, then the accretion would occur at approximately Bondi-Hoyle rate. When the feedback is present the picture is clearly more complicated, but for the argument's sake let us assume that the wind is strongly concentrated in some solid angle of fractional order unity and is localized along the polar axis of the disc. This seems to be in a qualitative agreement with observations of SS433, as is described in some detain in Section $7.7$ of  \cite{fab}. In this source, a range of polar angles $60^\circ < \theta < 90^\circ$ is being probed because of a considerable precession of the disc, and the velocity profile sharply rises from $V_w\simeq 100\hbox{km}/\hbox{s}$ for $\theta \simeq 90^\circ$ to $V_w\simeq 1500\hbox{km}/\hbox{s}$ for $\theta \simeq 60^\circ$. After positing that the wind is concentrated within some solid angle in the polar direction, we further assume that within this solid angle the accretion flow is disrupted  and the bow shock forms, while outside of this solid angle (i.e., closer the the disc plane) the accretion proceeds in the way not too different from conventional Bondi-Hoyle.

Let $\dot{M}$ be the rate at which the material is supplied to the disc. The angular momentum of the material accepted by the
disc is set by the density gradient $\nabla \rho$ of the ambient medium \citep{mcl}.
The momentum of the material accepted to the disc is transferred to the black hole at a rate 
\begin{equation}
    \vec{F}_{\rm acc}\sim-\dot{M}\vec{V}_*
    \label{facc}
\end{equation}
which is the frictional force due to accretion. The material is then accreted through the disc to some critical radius $R_{\rm cr}$, and mostly ejected
as fast and broad wind; the remaining material continues its way towards the black hole and the energy released during its accretion powers the outflow. The critical radius is probably related to the radius where the disc becomes quasi-spherical:
\begin{equation}
    R_{\rm cr}\sim {\dot{M}\over \dot{M}_{\rm Edd}}R_S
\end{equation}
where $\dot{M}_{\rm Edd}$ is the Eddington accretion rate and 
$R_s$ is the Schwarzschild radius of the black hole \citep{shak}. In this picture 
the velocity of the outflow is basically the escape velocity from this radius
\begin{equation}
    V_w=\chi \sqrt{\dot{M}_{\rm Edd}\over \dot{M}} c.
\end{equation}
The dimensionless number $\chi$ is estimated to be $\alpha/3$ to order of magnitude in \cite{shak},
which would give the outflow velocity of $\sim 1000(\alpha/3)\hbox{km}/\hbox{s}$ for SS433, clearly at least a factor of several smaller than the observed velocity.  The actual value of $\chi$ is yet to be confirmed by the numerical simulations, so perhaps a more sound approach is to approximately scale the wind velocity from the observed outflow in SS433:
\begin{equation}
    V_w\sim 1500\sqrt{1000\dot{M}_{\rm Edd}\over \dot{M}}{\hbox{km}\over \hbox{s}}.
    \label{vscaling}
\end{equation}

The fast wind will collide with the ambient gas along a bow shock, that will create an under-density downwind relative to the black hole. The computations from Section $2$ suggest that
if 
\begin{equation}
    V_W \gtrsim V_*,
    \label{vcriterion}
\end{equation}
    the gravitational force from the gas will be directed along $\vec{V}_*$ and be approximately given by 
Eq.~(\ref{eq:fric}):
\begin{equation}
    F_{\rm gr}\sim 2GM\sqrt{\dot{M}\rho V_w}V_*^{-1}.
    \label{gravity}
\end{equation}
The condition Eq.~(\ref{vcriterion}) implies 
\begin{equation}
    \rho\lesssim 3\times 10^{-10} \eta^{-1}\left({10M_{\odot}\over M}\right)\left({V_*\over 100\hbox{km}/\hbox{s}}\right)\hbox{g}/\hbox{cm}^3
    \label{rhocondition}
\end{equation}
where we have assumed that the disc is supplied at a fraction $\eta$ of the Bondi-Hoyle accretion rate
\begin{equation}
    \dot{M}=\eta \pi \rho (GM)^2 V_*^{-3}.
    \label{bondihoyle}
\end{equation}
The parameter $\eta$ can, even without the feedback, be substantially less than $1$ (MacLeod et al.~2017).

The equality between the two forces in Eqs.~(\ref{facc}) and (\ref{gravity}) is achieved at 
\begin{equation}
    \dot{M}\sim \dot{M}_B {V_w\over V_*},
    \label{criticalmdot}
\end{equation}
where $\dot{M}_B=\pi \rho (GM)^2/V_*^3$ is the Bondi-Hoyle accretion rate. If $\dot{M}$ is smaller than this value and Eq.~(\ref{vcriterion}) is satisfied, and
if our computation of the anti-friction force in Eq.~(\ref{gravity}) is correct, then  the overall force  is directed along the velocity $\vec{v}_*$ and the black hole accelerates. In fact Eq.~(\ref{vcriterion}) is already sufficient, since accretion physics guarantees $\dot{M}<\dot{M}_B$ and thus $\dot{M}$ is automatically smaller than the value in  Eq.~(\ref{criticalmdot}). 

We do not know what the realistic values of $\eta$ is when the feedback is included, however it is likely that the densities in Eq.~(\ref{rhocondition}) are applicable to the common envelope's outer edges (see e.g., the density profiles in Figure 1 of MacLeod et al.~2017). The NDF might well stop the inspiral at those small densities, and SS433 may well be the case in point. If however, the BH penetrates to deeper layers, then the wind velocity formally becomes smaller than velocity of the BH\footnote{Strictly speaking, the scalings in Eq~(\ref{vscaling}) is valid only for radiation-pressure dominated discs with Thompson opacity. However,
even for the very high accretion rates considered here, the discs are typically dominated by the radiation pressure at the spherization radius. The opacity is indeed likely to be higher than Thompson, which will in fact result at the spherization at larger radius and thus the slower winds}. In this case the wind is ineffective in creating an outflow and the gas does not have a chance to escape from the outer edge of the disc. As far as we know such a situation has not been previously explored in a systematic way. The closest study we are aware of is the low-angular-momentum ZEBRA solution of \cite{coughlin},
where the formal spherization radius is larger than the circularization radius of the disc. These authors argue that in that case, the gas instead of escaping, forms a weakly  bound atmosphere around the BH, and the built-up mass ensures that most of the material in the atmosphere makes its way to the immediate vicinity on the BH. One can then expect a significant fraction of the $\dot{M}c^2$ to come out as a collimated  relativistic outflow puncturing the atmosphere and dramatically affecting the environment far outside the Bondi radius. Whether this in fact happens, needs to be explored in numerical experiments, but obviously this scenario is very advantageous for NDF, as the collimated relativistic outflow is expected to inflate an under-dense bubble far outside the Bondi radius.

Inside AGN discs, the density is given by
\begin{equation}
    \rho=(3\pi\alpha)^{-1}{\dot{M}\over h^3\Omega},
    \label{rhoAGN}
\end{equation}
where $\dot{M}$ is the accretion rate through the disc, $h$ is the scaleheight, $\Omega$ is the angular frequency of the Keplerian motion, and $\alpha$ is the Shakura-Sunyaev disc viscosity parameters. To get a sense of typical numbers, we can rewrite the above as
\begin{eqnarray}
    \rho&\sim &3\times 10^{-12}(3\pi\alpha)^{-1}\hbox{g}/\hbox{cm}^3\\
     & &\times\left({10^8M_\odot\over M}\right)^{1/2}\left({r\over 0.1\hbox{pc}}\right)^{-3/2}\left({0.01\over h/r}\right)^3\nonumber
    \label{rhoAGN1}
\end{eqnarray}
Here $M$ is the mass of the supermassive black hole. The density is sensitive to the value $h/r$, which is difficult to
compute from first principles as the disc likely consists of turbulent multi-phased and possibly strongly magnetized gas. One useful benchmark is the critical density at which the gas becomes self-gravitating,
\begin{equation}
    \rho_{\rm cr}\sim M/r^3=7\times 10^{-12}\left({M\over 10^8M_\odot}\right)\left({0.1\hbox{pc}\over r}\right)^3\hbox{g}/\hbox{cm}^3
\end{equation}
A stellar mass black hole (or a neutron star)
 moving through an AGN disc, will likely accrete at a super-Eddington rate and may experience NDF. If the estimates above are representative of the discs' densities, then in some regions of the disc the accretion onto the compact objects will drive 
 winds that are faster then the motion of the object relative to the disc; this is the scenario for which we make a specific estimate for the NDF. Alternatively, for higher disc densities (e.g., closer to the accreting supermassive black hole) the winds from  the outer discs may be quenched and much more material will make it to the immediate vicinity of the compact object. The resulting relativistic jets would inflate a bubble that could also potentially result in NDF.  



The general scenario sketched above is full of uncertainties that must be explored through numerical experiments, the most obvious being how the outflow interacts with the inflow supplying  gas to the accretion disc around the black hole. We envisage two possible ways in which this may occur. The first possibility, sketched in Fig.~3, is that the outflow is channelled by the disc in such a way that the bow shock has 
a gap, through which the disc can be supplied. The second possibility 
is that the disc supply alternates in time with the outflow, creating a limit cycle. During the ``supply'' part of the outer edges of the disc increase in mass, while the inner parts are empty. As the material accretes inwards and reaches the photon trapping radius, the outflow is launched and the pressure from it shuts off the supply of the gas to the disc. The supply resumes when the inner parts of the disc become drained and the outflow stops. A limit cycle that seems to fit  the above description has been recently observed in numerical experiments of \cite{Regan}.

{Numerical experiments are also needed to clarify some of the physical elements that underpin our proposal.   It is now clear that the disk accretion of magnetized, poorly radiative gas leads to disk winds that divert the inflow \citep{bla}, so that the effective accretion rate becomes roughly proportional to disk radius \citep{beg12,yuan15}.  To predict how these winds interact with matter that has yet to fall onto the disk, one requires several additional pieces of information.  First, the angular distribution of the wind's velocity and mass flux will be needed.  Second, for disks that extend beyond the photon-trapping radius of steady accretion, it will be necessary to characterize whatever steady, unstable, or limit-cycle behavior results from the viscous and thermal instabilities in this regime \citep{abram88}, which may themselves be modified by wind emission \citep{janiuk02,shen14}.   It will be especially useful to explore each of these questions across a range of each important dimensionless parameter.  (Such parameters might include: the relative velocity or density difference across the black hole's Bondi-Hoyle radius; Bondi-Hoyle accretion rate in units of the Eddington rate; magnetization of the inflow; radiation-to-gas pressure ratio of the surrounding medium.)  }
\begin{figure}
\includegraphics[width=0.5\textwidth]{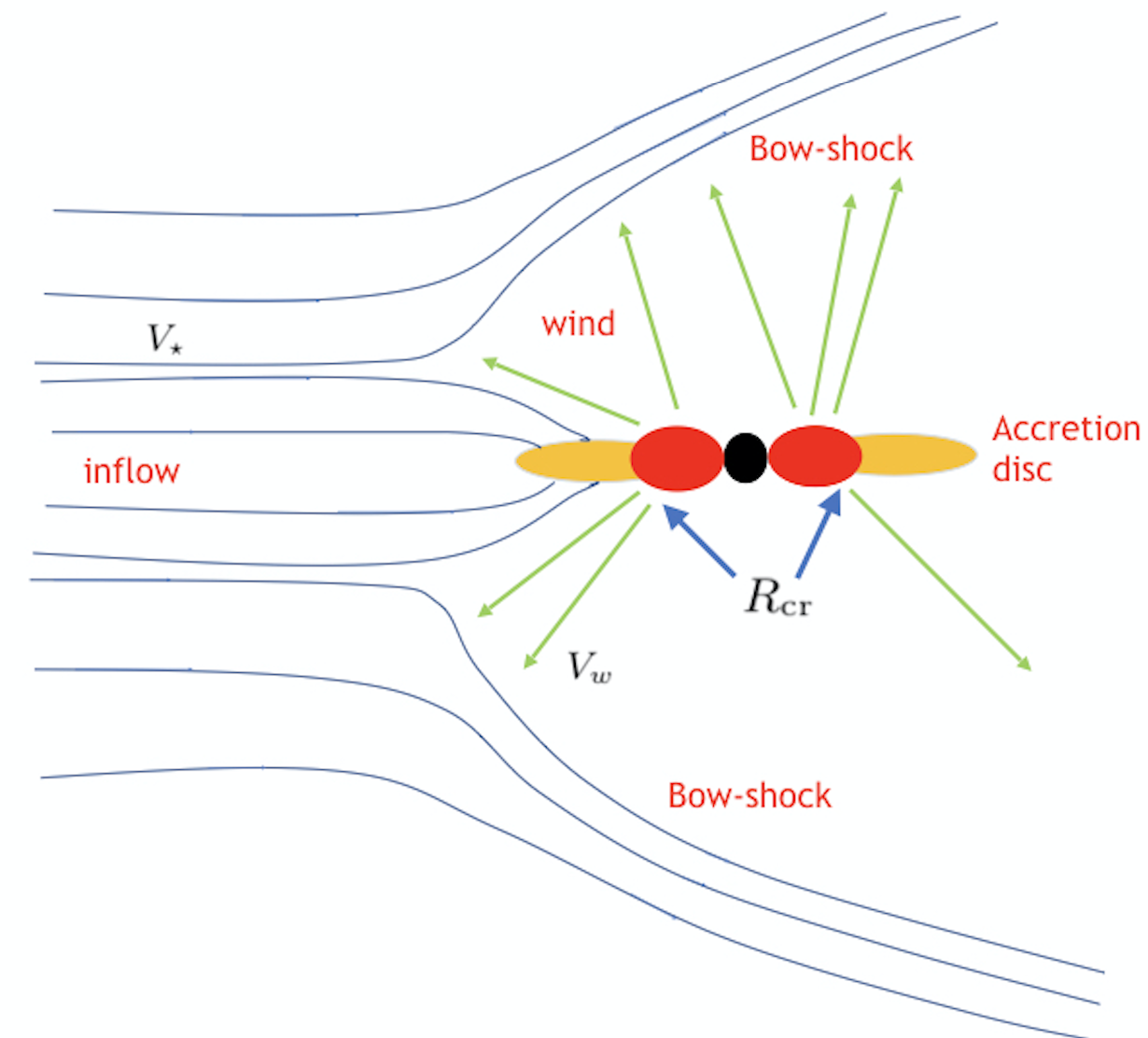}
\caption{Schematic view of the possible configuration of the accretion on a moving black hole. It is assumed that the angular momentum vector of the accretion disc is in the plane of the picture; this can be the case if e.g.~the density gradient is perpendicular to the page. The disc is supplied by the stream that is shielded by the disc from the wind. The wind itself originates from the inner part of the disc inside $R_{\rm cr}$.} 
\label{fig:bowshock2}
\end{figure}

\section{Possible consequences of Negative Dynamical Friction}
Here we allow ourselves to speculate briefly on the implications of the NDF. 

\subsection{``Grazing'' accretion for the black-hole inspiral through a stellar envelope}

The black hole entering the envelope of its companion will experience a period of highly super-Eddington accretion. It is far from clear however, whether the arguments of the 
previous section made for essentially an infinite gaseous medium, will carry over for the geometry of the common envelope. Nonetheless, if the friction inside the envelope is negative, the black hole will be pushed out to the outer layers of the envelope. The outflow will break through the surface layers and create a visible electromagnetic radiation across a wide range of frequencies, possibly similar to that seen in SS433. Following \cite{soker}, we refer to this mode of accretion as the ``grazing'' accretion, where a black hole fills its own Roche lobe 
by grazing the surface layers of the  companion, and then ejects most of it as a super-Eddington outflow\footnote{\cite{soker} argue that removal of the surface layers of the giant due to the feedback from accretion will by itself, without any appeal to the NDF, delay or prevent the onset of the formation of a common envelope}. One needs to investigate further the longevity and observational signatures of this mode of accretion.

\subsection{Stellar-mass black holes in AGN discs}
It has been argued that AGN discs may spawn formation of stellar-mass black holes inside them \citep{art,levin03,levin}.
These black holes will accrete the material from the disc, and simple estimates show that the Bondi-Hoyle accretion rate is strongly super-Eddington. The feedback from the accretion may change the direction of the migration of the black holes through the disc, potentially affecting their detectability as LISA sources. This is in analogy with the result found in \cite{velasco},
who found that  the migration of planetesimals with thermal feedback can be reversed. The NDF, if it operates in the disc, will also increase the inclination and eccentricity of the black hole orbit. The strongly accreting black hole will rise to the surface of the disc once per orbit, with the outflow from the black hole bursting through. This may contribute to AGN variability, which needs to be investigated in future work.

\subsection{Super-Eddington black hole growth in  early Universe}
There are reasons to believe that super-Eddington accretion onto intermediate-mass black holes may play a role in rapidly increasing their masses in the early Universe, and in producing the supermassive black holes currently observed in galactic nuclei (e.g., Madau \& Rees 2001). For this to occur, however,  both the black holes and dense clouds of gas feeding them must sink rapidly towards the lowest points of gravitational potentials of early halos, so dynamical friction is an essential ingredient of the story (e.g., Pfister et al.~2019 and references therein). On the surface it may seem that negative dynamical friction on black holes could be problematic for super-Eddington accretion. However, in the context of a dynamical friction in a galaxy, the purely gravitational dynamical friction due to scattering of stars is likely to play the dominant role. This is consistent with the findings of \cite{pfister}.
Still, it could be interesting to explore the effects of NDF driven by interactions with gas, if future numerical experiments demonstrate that NDF is indeed a likely outcome for a black hole accreting at a super-Eddington rate.

\section*{Acknowledgements}

We thank Konstantin Postnov, Juri Poutanen, and Sergei Fabrika for illuminating  discussions about SS433. We also thank Phil Armitage, Phil Chang, Xinyu Li, Morgan MacLeod, Hugo Pfister, Eliot Quataert, Enrico Ramirez-Ruiz, and Marta Volonteri for useful comments that helped to shape the astrophysical context of this work.
CDM's research is supported by an NSERC discovery grant.

\label{lastpage}
\end{document}